# A quantitative and a qualitative study of the resonance assisted double proton transfer in formic acid dimer.

Sharan Shetty[a], Sourav Pal[*, a], Dilip G. Kanhere[b], Annick Goursot[c]

[a]Theoretical Chemistry Group, Physical Chemistry Divison, National Chemical Laboratory, Pune-411008, India.
[b]Centre for Modeling and Simulation and Department of Physics, University of Pune, Pune-411008, India
[c]Ecole de Chimie de Montpellier, UMR 5618 CNRS Ecole de Chimie, 8, rue de l'Ecole Normale 34296 Montpellier, Cedex 5, France

**Abstract:**

We have performed *ab initio* molecular dynamics simulations to study the nature of the synchronous double proton transfer in formic acid dimer. In order to understand the evolution of the bonding during the double proton transfer, we have used the electron localization function and the molecular orbital isosurfaces. During the dynamics of the double proton transfer in formic acid dimer the two formic acid monomers approach each other, forming Speakman-Hadzi type, of short strong hydrogen bonds at the transition state. The Speakman-Hadzi type of short strong hydrogen bonds are also shown to be polar covalent bonds. Based on the concept of resonance assisted hydrogen bond proposed by Gilli *et al* (Gilli P, Bertolasi V, Ferretti V & G. Gilli, *J. Am. Chem. Soc.*, 116 (1994). 909-915), we show that the proton at the transition state is connected by a π-conjugated O—C—O structures, indicating a resonance assisted hydrogen bond. On this basis the double proton transfer process in the formic acid dimer can be termed as a resonance assisted double proton transfer. Our finding on the above analysis suggest that the synchronous double proton transfer in the formic acid dimer is due to the formation of (a) Speakman-Hadzi type of short strong hydrogen bonds and (b) Resonance assisted hydrogen bond at the transition state.

## I. INTRODUCTION

For the past several years inter and intra-molecular proton transfer (PT) in hydrogen bonded systems has been recognized as one of the most fundamental and important phenomena in chemical and biological sciences.[1-14] Thus, PT has received considerable attention, both from theoretical and experimental point of view.[2,3] Single proton transfer (SPT) has been extensively studied in the last few decades, where malonaldehyde has been the prototypical molecule, particularly for the study of intra-molecular SPT.[4,5] Multiple proton transfer (MPT), which involves more than one PT, plays a crucial role in bio-chemical and organic processes.[6] Double proton transfer (DPT) is one of the model examples of MPT which involves the transfer of protons either synchronously or asynchronously in intermolecular hydrogen bonded systems.[7,8] Tautomeric proton transfer in the hydrogen bonded base pair of DNA and proton relay systems in many enzymatic reactions are typical examples in biological systems.

Interestingly, recent studies have shown that the excited state dynamics of PT gives rise to significant quantum mechanical effects, such as tunneling.[9,10] In a recent study, tunneling was shown to play a significant role in the PT even at room temperature.[11] Recently, Ushiyama and Takatsuka studied the time dependent probability of quantum tunneling in tropolene using a quasisemiclassical approach.[12] Moreover, it has also been shown that the solvent effects in the PT reactions can change the nature of PT.[10(c), 13]

It has been shown that during the PT there is formation of short strong hydrogen bonds (SSHB) or low barrier hydrogen bonds (LBHB) which play important role in stabilizing the transition state in many enzymatic reactions.[14-16] Frey et al found the evidence of LBHB in the catalytic triad of a serine protease between aspartate and histidinium.[15(b)] Kim et al recently showed that the SSHB involved in the formamidine-formic acid dimer is of Coulomb type where the H atom is more localized,[16] rather than a Speakman-Hadzi type of in which the H atom is delocalized between the two homonuclear atoms.[14,17] Interestingly, in the last decade Gilli et al used the crystal structure correlation method to categorize the SSHB into resonance assisted hydrogen bonding (RAHB) and charge assisted hydrogen bonding (CAHB).[18,19] They showed that in RAHB the H atom is attached to the two oxygen atoms through π-conjugated double bonds. Recently, ab-initio path integral techniques have been implemented to study the proton transfer properties of the charged water complexes where the strength of the hydrogen bond has been discussed at room temperature.[5] Hence it is interesting to understand that how the SSHB's are formed during the PT reaction and how do they affect the PT process.

It is worth mentioning that during the transfer of proton from one atom to another, the surrounding atoms should adjust to new equilibrium positions. These environmental effects due to the PT processes are accompanied by a significant reorganization in the charge distribution.[20] This electronic rearrangement during the PT may lead to spectral changes that can be very important in photochromic molecules.[3] Although the PT reactions involve the change in the equilibrium positions of the surrounding nuclei, thus affecting the electronic charge distribution, there has been much less discussion on the change in the nature of bonding during the PT. Recently, Krokidis et al.[21] used electron localization function (ELF) and catastrophe theory to discuss the change in the bonding pattern of C=C and C—C=O pi conjugation during the SPT in malonaldehyde and showed that these heavy atoms have a non-vanishing effect on the PT rate. In a very recent work Chocholoušová et al.[22] explained the red-shifted O—H----O and improper blue shifted C—H----O hydrogen bonds of the first local minimum of the formic acid dimer (FAD) using natural bond orbital analysis and rehybridization. Interestingly they showed that the red shifted O—H----O is due to the hyperconjugation and improper blue shifted C—H----O has a mesomeric structure which is due to the delocalization of electron density from C—H $\sigma^*$ antibonding orbital.[22]

In the present study, we have chosen FAD as a model system to study the evolution of bonding during the DPT. To study the bonding analysis, we have used ELF and molecular orbital pictures in combination with the *ab initio* molecular dynamics (AIMD) simulation in a canonical (constant temperature) ensemble.

FAD is considered to be a strong hydrogen bonded (O-----O < 2.7 Å) system.[23] Activation barrier and the structural properties of FAD have been well studied.[8, 23-27] Due to the low mass of protons it has been argued, whether the DPT process in FAD is a quantum tunneling or a classical effect.[8(b), 23-27] Madeja and Havenith in a recent experimental work used first fully analyzed high resolution spectrum to show that the tunneling splitting of the DPT in DCOOH dimer is 0.00286 cm$^{-1}$, corresponding to a proton transfer time of 5.8 ns.[24] In this study the energy barrier was shown to be 12.1 kcal/mol. Madeja et al.[28] have also identified polar isomers of formic acid dimers in helium droplets. Kim used variational transition state theory (VTST) using multidimensional semiclassical tunneling approximations to show that the tunneling is effective in FAD.[23] He calculated the barrier height of 8.94 kcal/mol at a G2* level. Chang et al.[8(a)] obtained a tunneling splitting of 0.3 cm$^{-1}$ (proton transfer time ~ 55ps). Shida et al.[8(b)] used a 3D potential energy surface and obtained a barrier height of 12.0 kcal/mol and a tunneling splitting of 0.004cm$^{-1}$. Vener et al.[25] used the same approach as Shida et al but with a more accurate *ab initio* calculations showed the tunneling splitting to be 0.3 cm$^{-1}$. In a very recent work, Tautermann et al[27] used the VTST including the quantum chemical effects like tunneling to calculate the reaction rates and the transmission coefficients in the temperature range of 100 to 400 K and for various isotopic substitutions. Ito and Nakanaga have shown that the spectral pattern for $(HCOOH)_2$ and $(DCOOH)_2$ isotopomers are different from each other.[29] Later, Florio et al.[30] showed that the O—H and the C—H stretch of local modes of FAD mix while forming the normal modes. In an interesting experiment, Howard and Kjaregard used an intracavity laser photoacoustic spectroscopy to study the vibrational analysis of the formic acid isotopomers HCOOH and DCOOH.[31] They observed that the O-H stretching vibration in HCOOH and DCOOH is quite different. In an another spectroscopic experiment, Matylitsky et al.[32] for the first time found the rotational constants of the regular O—H----O/O----H—O isomer of $(HCOOH)_2$. From the above spectroscopic study, it is clear that the O—H frequency in $(DCOOH)_2$ and $(HCOOH)_2$ are different. Hence it is questionable to compare the results of the DPT in $(DCOOH)_2$ and $(HCOOH)_2$.

Moreover, earlier studies on FAD have confirmed that there is a synchronous DPT in the gas phase FAD.[8, 23-27, 33-35] In an important work, Miura et al.[34] have applied the AIMD technique to study the synchronous DPT in FAD for a period of ~5ps. Ushiyama et al.[35] also used the same approach as Miura et al[34], but found the proton transfer time of 150 fs, which is too low compared to the earlier studies. Wolf et al performed molecular dynamics using a projected augmented wave method to study the simultaneous proton transfer in FAD.[33] They carried the dynamics for a period of 20 ps at a temp of 500-700 K. However, asynchronous DPT was shown to take place in the

presence of the solvent by Kohanoff *et al*.[36]

On the earlier background, we revisit the dynamics and the structural investigation of DPT in FAD by using the AIMD simulation in a canonical ensemble (constant temperature). Further, we focus our investigation to understand the change in the bonding of FAD during the DPT.

We have used ELF and molecular orbital isosurfaces to investigate the evolution of bonding during the DPT process in FAD. Becke and Edgecombe proposed the concept of ELF which is one of the important tools to understand the nature of bonding in many chemical systems.[37, 38] ELF has been used to elucidate sigma, pi, polar covalent, ionic bonds and the delocalization in the aromatic systems and in metallic bonding.[38, 39] ELF also accounts for the change in the bonding during a chemical reaction.[39]

The paper has been organized as follows: In section II the technical details of the DPT in FAD using the AIMD simulations has been discussed and also a brief discussion on ELF is given. In subsection III (A) a comparison with the earlier study on DPT has been made. Further, the formation of SSHB and resonance assisted DPT has been discussed. In subsection III (B) evolution of bonding during the DPT has been illustrated. Conclusions are presented in section IV.

## II. THEORETICAL METHODS AND COMPUTATIONAL DETAILS

All the calculations presented in this paper have been performed by using VASP code.[40] The AIMD technique used in VASP[41] treats the nuclear motion classically and simultaneously evaluates the instantaneous electronic ground state using the Kohn-Sham equations within the generalized gradient approximation (GGA) with the Perdew-Wang 91 (PW-91) functional.[42] We have used GGA instead of local density approximation (LDA), since GGA has been shown to give an accurate description of hydrogen bonds.[43] The present method uses the plane wave basis set in conjunction with the ultra-soft pseudopotentials.[44] Miura *et al*.[34] have already shown that the results obtained for the DPT in FAD from the localized basis set and the plane waves were same. The dimer was placed in an orthorhombic supercell of dimension a=13.19Å , b=4.00 Å , c=11.59Å. The equations of motion were integrated with a time step of 0.5 fs. The forces on the atoms after each dynamic step were converged to ~0.001 eV/Å. Earlier theoretical work have carried out the study of the DPT in FAD in the range of 100 to 400 K.[27,23,33,34] In the present study we keep the temperature of the FAD system at 200±40 K using Nose-Hoover thermostat.[45] The total simulation was carried out for a period of 25 ps. Initially the FAD system was thermally equilibrated for a period of 2.5 ps, where we observe large fluctuation in the temperature. A constrained dynamics was also performed to understand the effect of the motion of the heavy atoms.

The nature of bonding was studied using ELF. There are several reviews, which discuss the technical details and applications of ELF in a broader perspective. Here, we give only a brief discussion on ELF. In the present study, ELF is calculated as

$$ELF = \frac{1}{1+\left(\frac{D}{D_h}\right)^2}$$

$$D = \frac{1}{2}\sum_i |\nabla \varphi_i|^2 - \frac{1}{8}\frac{|\nabla \rho|^2}{\rho}$$

$$D_h = \frac{3}{10}(3\pi^2)^{\frac{5}{3}} \rho^{\frac{5}{3}}$$

where $\varphi_i$ are the Kohn-Sham orbitals. D is the excess local kinetic energy due to the Pauli repulsion obtained from the difference between the definite positive kinetic energy density of actual fermioinic system and that of the von Weiszacker kinetic energy functional and $D_h$ is the Fermi constant. It has been shown that the ELF values lie between 0 and 1.[37] The excess kinetic energy takes low values in regions occupied by the paired electrons and this leads to the values of the ELF close to unity showing a localized domain, while low values of ELF reflect a delocalized region as in metallic systems.[38] It is, therefore possible to define regions of space that are associated with different electron pairs in a molecular system. The present ELF calculations are based on the valence electrons, while the core electrons are replaced by the ultra-soft pseudopotential.

## III. RESULTS AND DISCUSSION

We divide this section in two parts. In the first part, we present the structural and dynamical aspects of the DPT and in the later part we do a critical analysis of the evolution of the bonding during the DPT in FAD.

**(A) Structure and Dynamics**

The structural parameters for the equilibrium and the transition state structures of FAD are given in Table. I. The DPT process in FAD is schematically described in Scheme 1. Structure (1) and (3) in Scheme 1 describes the two equilibrium structures, while (2) represents the transition state structure on the potential energy surface.

Bond length distribution of the covalent bond (r1) and the electrostatic hydrogen bond (r2) for the total simulation period from 5ps to 25 ps (equilibration period is not included) is given in Fig. 1, where r1, r2, r3 and r4 are the internal coordinates defined in Scheme 1(1). For the ~16 ps the O—H (r1) and the O----H (r2) bonds fluctuate in there mean positions and no PT is observed (Fig. 1). Around 18.8 ps the r1 bond lengthens and the r2 bond shortens. At ~18.8 ps both the bonds cross each other and the r1 bond is converted into r2 and vice versa (shown by an arrow), hence indicating a PT. In the same manner the other proton in FAD also simultaneously moves from one monomer to another, when r3 changes to r4 and vice-versa and has a similar behavior as shown in the Fig. 1. This reveals that there is a DPT taking place in FAD at ~18.8 ps. This confirms the earlier theoretical investigations where the DPT in FAD was shown to take place in the ps range.[8(a), 25, 33, 34]

Earlier studies have successfully analyzed the synchronous nature of DPT in FAD.[8, 33-35] Before we proceed to the analysis of O—C—O structure it is necessary to give a brief discussion on the synchronous DPT process in FAD. From now onwards we present the data between 17 ps and 20 ps where the evolution of the DPT process takes place as shown in Scheme. 1. To illustrate the simultaneous motion of both the protons in FAD, we plot the change in the bond length distribution of the covalent O—H bonds (r1 versus r3) (Fig. 2) and the electrostatic O----H hydrogen bonds (r2 versus r4) (Fig. 3).

The linear regression of the graphs in Fig. 2 and Fig. 3 have the same slope of 0.98, hence implying that the change in the bond length of covalent O—H (r1 and r3) bonds and the hydrogen O----H bonds (r2 and r4) is symmetric. This shows, that the motion of both the protons from one monomer to the other is synchronous, which agrees with the earlier results.[23-27, 33-35] It is noteworthy to mention that if the slopes of the graphs in Fig. 2 or Fig. 3 deviate from value 1 the proton motion during the DPT becomes asynchronous. Thus, the slopes of the graphs of Fig.2 and Fig. 3 together decide the degree of synchronization of the DPT process.

Moreover it has been shown that the DPT process in FAD is initiated by a concerted inter-monomer vibration, which brings both the monomer units close to each other.[8,16] This allows the O-----O distance to reduce and favors a synchronous PT. On this basis we study the dynamics of the O-----O distance which indicates the inter-monomer change and correlate it with the O—H and O----H bond length fluctuations. To discuss this issue we define two parameters

$$\rho1 = (r2-r1) + (r4-r3),$$
$$\rho2 = r1+r2+r3+r4$$

$\rho1$ indicates the O—H and O----H bond length fluctuations and the change in $\rho2$ indicates the inter-monomer fluctuations during the DPT between 17 to 20 ps.

The plot of $\rho1$ versus $\rho2$ is given in Fig. 4. Interestingly, we see that the quadratic regression of the graph of $\rho1$ versus $\rho2$ (Fig. 4) is a parabola. The graph in Fig. 4 agrees very well with the earlier results by Miura *et al.*[34] Kim *et al* showed that the intermonomer vibration during the DPT in the formaidine-formic acid dimer had a similar behavior as in Fig. 4, however the nature of DPT was shown to be asynchronous.[16] The positive and the negative values of $\rho1$ in Fig. 4 indicate the fluctuations in the O—H and O----H bonds of FAD before and after the DPT, respectively. The structure of FAD at $\rho1=0$ (where r1=r2=r3=r4) is the transition state (Scheme 1 (2)) which is observed at ~18.8 ps. The barrier height for the DPT, which is 6.6 kcal/mol is given in Table II.

We can also see from Table II that the DPT barrier calculated by us using PW-91/plane wave and the other higher level theoretical calculations match within ~1-2 kcal/mol. It should be noted that the, barrier for the DPT in FAD obtained in the present investigation and by most of the earlier theoretical calculations disagree with the experimental results for the DPT in DCOOH dimer (Table. II). Moreover, the time scale for the DPT in DCOOH was shown to be in nano-second[28] which does not match with some of the earlier theoretical results.[33-35] The reason for this would either be the deuterium effect in DCOOH dimer as discussed earlier or the higher tunneling splitting of the DPT in DCOOH dimer or maybe both.[28] It is out of scope to discuss this issue in the present investigation. However, we focus on the change in the bonding structure during the DPT in FAD.

The minimum of the parabola (Fig. 4) confirms that the transition state has the shortest inter-monomer distance, where the O-----O distance is 2.468 Å (Table. I). At the transition state ($\rho1=0$), the H atom lies between the oxygen atoms. This allows the H atom to get delocalized between the two oxygen atoms, hence forming a Speakman-Hadzi type of SSHB.[14, 17] The shortening of the intermonomer distance and the formation of Speakman –Hadzi type of SSHB stabilizes the transition state and the barrier is lowered. This lowering of the transition state barrier due to the formation of SSHB, may allow the protons to cross the barrier classically rather than a quantum tunneling. The formation of SSHB also makes the H atoms to form polar covalent bonds with the oxygen atoms on the either side, which is discussed in the next section. In the initial phase of the simulation we observe an asynchronous motion of the protons which may be due to larger intermonomer distance resulting in weak hydrogen bonds. It is evident from the above discussion that, the DPT takes place when the intermonomer distance is at minimum and result in the formation of Speakman-Hadzi type of SSHB.

Ushiyama *et al*.[35] showed that the deformation in the O—C—O structure in FAD plays a vital role during the DPT. Towards this, we focus our study on the bond length fluctuations of C=O and C—O. Gilli *et al* have used λ parameter to analyze the RAHB in many hydrogen bonded systems.[18, 19] In the present work we generalize the λ parameter to discuss the resonance in the O—C—O structures. We define two λ parameters corresponding to the two O—C—O structures in FAD,

$$\lambda1 = \{1 + Q1/Q^0\}/2.0$$
$$\lambda2 = \{1 + Q2/Q^0\}/2.0$$

Where, $Q_1 = r_6-r_5$ and $Q_2 = r_7-r_8$, where $r_5$, $r_6$, $r_7$ and $r_8$ are shown in Scheme 1(1), and $Q^0$ is the same quantity for the pure single (C—O) and double bond (C=O) which remains constant. In our study $Q^0 = 0.07$. When $Q_1 = Q_2 = 0$ the bonds between C and O atoms are equal and $\lambda_1 = \lambda_2 = 0.5$, which indicates a resonance within O—C—O structures. We plot $\lambda_1$, $\lambda_2$ as a function of time between 17-20 ps which corresponds to the DPT process in Fig. 5. In Fig. 5 before the DPT, $\lambda_1 = 1$ and $\lambda_2 = 0$ indicates the bond length fluctuations of C—O and C=O of the structure (1) in Scheme 1. At $\lambda_1 = \lambda_2 = 0.5$, where the horizontal line cuts the graph (shown by an arrow in Fig. 5) indicates the resonance in the O—C—O structures. We can also observe that the horizontal line cuts the graph at ~18.8 ps where the transition state was located (Fig. 4). $\lambda_1 = 0$ and $\lambda_2 = 1$ indicates the bond length fluctuations of C—O and C=O of the structure (3) in Scheme 1. Hence the above observation reveals that there is resonance in both the O—C—O structures of FAD during the DPT.

Next we establish the correlation of the resonance in the O—C—O structures and the dynamics of the synchronous DPT. For this we plot $\rho_1$ versus $\rho_3$ (Fig. 6), where $\rho_1$ is already defined earlier and $\rho_3 = (r_8-r_7) + (r_6-r_5)$. $\rho_1$ is the fluctuations in all the four O and H bonds, $\rho_3$, show the dynamical change of the C and O bonds. The correlation coefficient of the linear regression of this graph is 0.94. This shows that there is a simultaneous evolution of the resonance in the O—C—O structures and the synchronous motion of the DPT. We can also see that the linear regression passes through $\rho_1=\rho_3=0$, hence indicating that the resonance in the O—C—O structures occurs at the transition state. This reveals that, the H atoms at the transition state which are Speakman-Hadzi type of SSHB, are attached to the oxygen atoms on the either side by a π-conjugated double bonds and can be termed as RAHB. Hence our finding on the structural and dynamical data of DPT in FAD clearly demonstrates that the formation of Speakman-Hadzi type of SSHB and the RAHB leads to synchronous DPT in FAD. Hence the DPT in FAD is a resonance assisted DPT.

In order to investigate the role of heavy atoms, we have also carried out a constrained dynamics simulation by fixing all the atoms except the hydrogen atoms involved in the hydrogen bonding. We found that the hydrogen atoms only wiggle at the same position and no proton transfer was observed. It can be seen that, for the proton transfer to take place it is necessary to relax all the degrees of freedom in the system.

(**B**) **Bonding**

In the present section, we discuss the bonding evolution during the DPT in FAD with the help of ELF at different isosurface values and the molecular orbital at one third of the maximum isosurface value.

Fig. 7 describes the ELF isosurfaces of the ground state (left column), isomer (right column) and the transition sate (middle column) of FAD at different isosurface values. Fig. 8 describes the molecular orbital isosurfaces of the ground state structures (left and right) and the transition state (middle) at one third of its maximum isovalue. Five different localization regions have been recognized as shown in Fig.7, which change during the proton transfer in FAD.

ELF=0.78 isovalue (Fig. 7 (a)) of the ground state FAD shows that there is a circular electron localization, indicated by region I which describes a double bond between the C and O atoms. Regions II and III show **sigma** bond between C—O and O—H respectively. No localization region is seen in the region IV, this is due to the existence of the hydrogen bond between the oxygen of one monomer of FAD with the hydrogen of the other. Region V represents the lone pair of electrons of the oxygen atoms. At a higher isosurface value of ELF = 0.8 (Fig. 7 (b)) all the localization regions are completely separated and clearly indicate double bond, single bond and the lone pair of electrons. Even at a higher ELF isovalue of 0.84 (Fig. 7 (c)) we see localization in the regions III and V, which imply that the lone pairs and the covalent O—H bonds are highly localized. Fig. 8 shows the isosurface of the highest occupied molecular orbital (HOMO) which shows a pi-orbital of the O of the C=O.

However, the ELF isosurfaces of the transition states are strikingly different than the ground state. ELF at 0.78 (Fig. 7(d)) of the transition state shows a symmetric electron localiztion isosurface. The localization in the regions I and II are same on both the monomers of FAD. This indicates that all the C—O bonds are same. Moreover the region V also shows an identical localization domain on all the oxygen atoms. At ELF = 0.8 (Fig. 7(e)) we clearly see that all the four C—O bonds have similar localization. This reveals that there is a resonance in the O—C—O region of both the monomers. This is in agreement with the earlier discussion on structure and dynamics where we made a point that, at the transition state all the four C—O bonds are equal. However we also see a localization domain in the region IV which was seen to be empty in the ground state (Fig. 7(a)). Surprisingly, we don't see any localization region on the H atoms involved in the hydrogen bonds indicating $H^+$ ions. We also see that the region V, which was shown to be a lone pair of electrons in the ground state (Fig. 7(a)) has been deformed and the localization region is polarized towards the $H^+$ ions. If we increase the ELF value from 0.8 to 0.84 isovalue (Fig. 7(f)) we observe a symmetric localization region between the O—H and H—O bonds. This is due to the protons lying in the mid-way of O----O, which has been discussed in the earlier section. Interestingly, this arrangement of the $H^+$ lying in the center of the oxygen atoms and polarizing the electron density over the oxygen atoms allows it to form an intermediate electrostatic-covalent bond with the oxygen atoms on the either side (Fig. 7(f)), which is often called as a polar covalent bond.[38(b)] This kind of polar covalent bond has been seen in some enzymes when the H lies in the middle of the two oxygen atoms.[46] The formation of polar covalent bonds at the transition state can be attributed to the formation of Speakman-Hadzi type of SSHB. The isosurface of HOMO (Fig. 8(b)) of the transition state shows a pi orbital on all the oxygen atoms which is in contrast with the HOMO of the ground state (Fig. 8(a)) where the pi orbitals were seen only on the carbonyl oxygen atoms.

In Fig. 7 the right column shows the ELF isosurfaces of the isomer of the ground state of FAD after both the proton transfers take place from one monomer to the other through the transition state. From Fig. 7 we clearly see that the ELF isosurfaces of the ground state FAD (left column) and its isomer (right column) are mirror images of each other. The regions I, II, III of the ground state (Fig. 7) have been changed to II, I, IV of the isomer (Fig. 7(g)) respectively. The HOMO isodensity (Fig. 8 (c)) shows pi orbitals on the oxygen atoms of C=O as expected.

We show from the above study that ELF along with the HOMO isosurfaces play a significant role in understanding the evolution of the bonding structure during the DPT in FAD.

## IV. Conclusion

We have performed AIMD simulations for a period of 25 ps at 200±40 K in combination with a qualitative analysis of the ELF and molecular orbital analysis to investigate the nature of the synchronous proton transfer of DPT in FAD. Our investigation show that the synchronous motion of the protons during the DPT in FAD is due to the two important factors *viz.,* (a) Heavy atom (oxygen atoms) displacement which allows the formic acid monomers to approach each other and causes the H atoms to delocalize between the oxygen atom, forming a Speakman-Hadzi type of SSHB, which is also a polar covalent bond. (2) Resonance in both the O—C—O structures of FAD allows the H atom to form RAHB. Hence on the account of the RAHB, the DPT in FAD can be termed as resonance assisted double proton transfer. Our study suggests that if the H atom is involved in RAHB and a Speakman-Hadzi type of SSHB the proton transfer is synchronous. However, further theoretical as well as experimental studies are necessary to know whether the DPT in FAD is a classical effect or a quantum tunneling. Further theoretical investigations are necessary to understand if the discrepancies between the theory and experiments are due to the deuterium effect or any fundamental lacuna in the conventional molecular dynamics calculations used. The present theoretical investigation however, should suffice to understand the bonding structure during the PT reactions, which would be otherwise difficult to understand through experimental methods. The present work would also help in understanding the role of SSHB and RAHB during the PT reactions in biological systems such as DNA base pairs and enzymatic reactions.

**Acknowledgements**

S. Shetty, S. Pal and A. Goursot gratefully acknowledge the Indo-French Center for the Promotion of Advance Research (IFCPAR) (Project. No. 2605-2), New Delhi, India, for financial assistance.

*To whom correspondence should be addressed: Email: `pal@ems.ncl.res.in`

**TABLE. I** Structural parameters of the equilibrium and the transition state structures obtained us from *ab initio* molecular dynamics simulations using Parr-Wang 91(PW-91) functional have been compared with the experimental, and the density functional using Becke Lee Yang Parr exchange correlation functional. Bond lengths are given in Å, bond angles in degrees.

|  | Exp.[a] | Equilibrium structure | | Transition state | |
|---|---|---|---|---|---|
|  |  | AIMD (PW-91) | B-LYP[b] | AIMD (PW-91) | B-LYP[b] |
| d(O⋯O) | 2.703 | 2.611 | 2.645 | 2.468 | 2.441 |
| d(H⋯O) | 1.667 | 1.570 | 1.626 | 1.236 | 1.220 |
| d(O-H) | 1.036 | 1.033 | 1.019 | 1.232 | 1.220 |
| d(C-O) | 1.323 | 1.308 | 1.322 | 1.280 | 1.272 |
| d(C=O) | 1.220 | 1.238 | 1.233 | 1.282 | 1.272 |
| d(C-H) | 1.082 | 1.102 | 1.098 | 1.080 | 1.097 |
| ∠O-H⋯O | 180.0 | 178.8 | 179.9 | 176.4 | 177.6 |
| ∠H⋯O=C | 125.3 | 120.4 | 122.5 | 114.4 | 115.5 |
| ∠C-O-H | 108.2 | 112.0 | 110.6 | 117.4 | 115.5 |
| ∠O=C-O | 126.2 | 126.6 | 126.4 | 126.8 | 126.7 |
| ∠H-C=O | 115.4 | 120.7 | 121.7 | 118.5 | 116.6 |

[a]reference 48
[b]reference 30

**TABLE. II** The barrier height of the double proton transfer in formic acid dimer calculated by Parr-Wang-91 (PW-91) presented in this work is compared with the other theoretical calculations.

| Method/Basis set | Barrier height (kcal/mol) | References |
|---|---|---|
| Expirement[a] | 12.10 | 24 |
| PW-91/Plane wave | 6.60 | Present work |
| B-LYP/Plane wave | 5.40 | 30 |
| B3-LYP/6-31G(d,p) | 5.40 | 47 |
| B3-LYP/6-311G(2d,2p) | 6.40 | 49 |
| CCSD(T)/ag-cc-pVTZ//MP2/aug-cc-pVTZ | 7.89 | 27 |
| G2[*] | 8.94 | 23 |
| MP2/aug-cc-pVTZ | 6.71 | 27 |

[a]The experimental barrier given is for $(DCOOH)_2$.

**Figure Captions:**

Scheme. 1: Schematic representation of the double proton transfer in formic acid dimer. The structure (1) and (3) represent the ground state and its isomer. Structure (2) represents the transition state.

Figure. 1: The bond length fluctuation of covalent bond O—H (r1) shown by black line and hydrogen bond O---H (r2) shown by the red line for a period of 25 picosecond at T = 200±40 K. The arrow indicates the conversion of r1 to r2 and vice-versa.

Figure. 2: Bond length fluctuation of the covalent O—H bonds of the formic acid dimer r1 and r3 in the range of 17 picosecond to 20 picosecond. The linear regression is shown by a straight line.

Figure. 3: Bond length fluctuation of the hydrogen bonds in the formic acid dimer r2 and r4 in the range of 17 picosecond to 20 picosecond. The linear regression is shown by a straight line.

Figure. 4: The distribution function for the reaction coordinate $\rho 1$ versus $\rho 2$ of the formic acid dimer in the range of 17 picosecond to 20 picosecond.

Figure. 5: Plot of the coupling parameter $\lambda 1$ (black line), $\lambda 2$ (red line) as a function of time between 17 picosecond and 20 picosecond. The arrow indicates $\lambda 1=\lambda 2=0.5$ showing a resonance in both the O—C—O structures of the formic acid dimer.

Figure. 6: Plot of the distribution function of $\rho 3$ as a function of $\rho 1$ in the range of 17 picosecond to 20 picosecond of the formic acid dimer.

Figure. 7: Left column: 3D representation of the ELF of the ground state structure of Scheme. 1 (1) at different isovalues. The black spheres are the C atoms, the white spheres indicate the H atoms the grey spheres represnt the O atoms. (a) ELF=0.78, (b) ELF=0.82 (c) ELF=0.84. The regions I and II represent the C=O and the C—O respectively, region III represent the O—H bond, region IV represent the hydrogen bond between the H of one monomer with the O atom of the other, region V shows the lone pair of electrons on the oxygen atom of C=O.
Middle column: 3D representation of the ELF of the transition state structure shown in Scheme 1(3). (d) ELF=0.78, (e) ELF=0.82 (f) ELF=0.84. The ELF distribution shows a symmetric distribution. Region I and II indicates the resonance in the O—C—O structure. Region IV indicates the polar covalent bond between the O and H atoms.
Right column: 3D representation of the ELF of the isomer structure of Scheme 1 (3) at different isovalues. (g) ELF=0.78, (h) ELF=0.82 (i) ELF=0.84. The ELF isosurfaces of the isomer are the mirror images of the ground state structure (left column).

Figure. 8: 3D isosurfaces of the highest occupied molecular orbital (HOMO) at one third of the maximum isovalue of the (a) ground state, (b) transition state and (c) isomer. The black spheres are the C atoms, the white spheres indicate the H atoms the grey spheres

represent the O atoms. The isosurface of the HOMO of the ground state (a) and the isomer (b) show the pi orbital on the O atoms of the C=O. The isosurface of HOMO of the transition state (c) shows the pi-orbial on all the oxygen atoms.

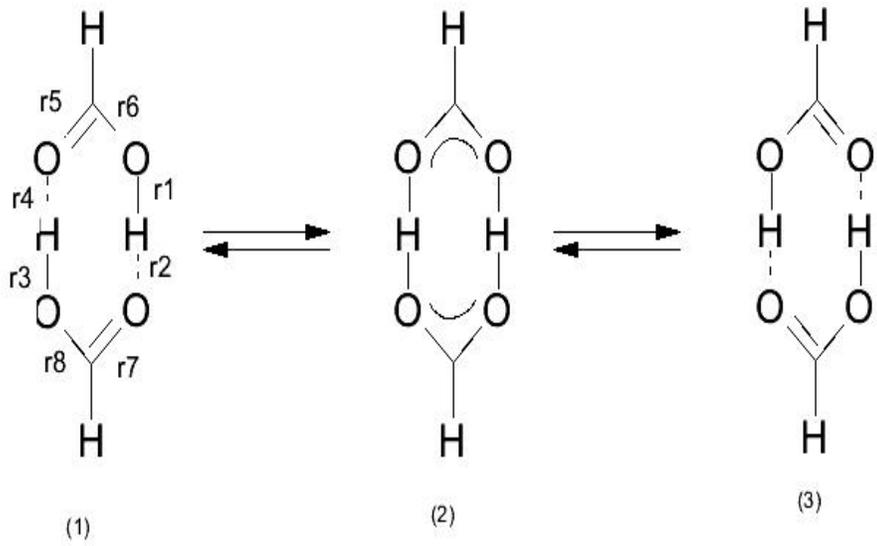

Scheme. 1

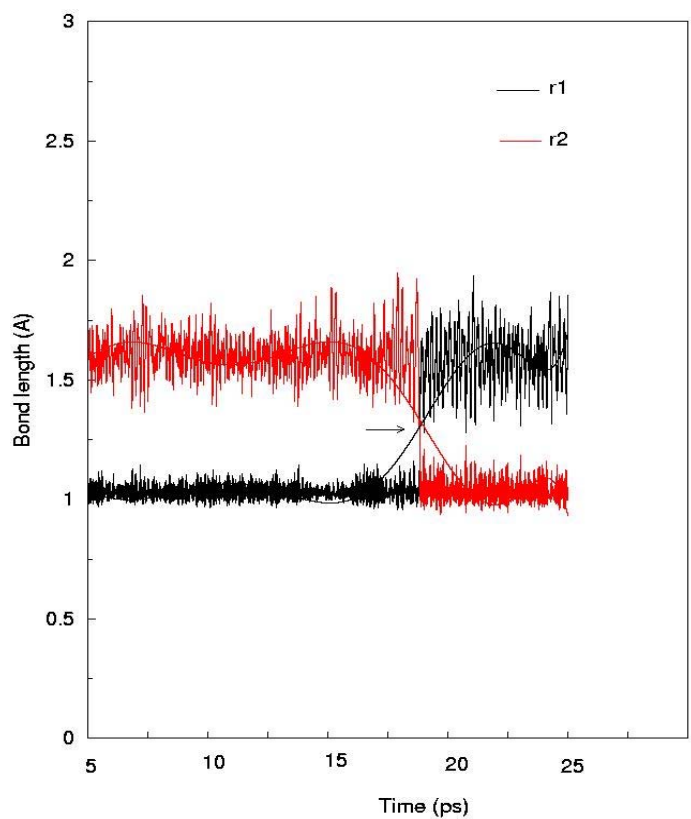

Fig. 1

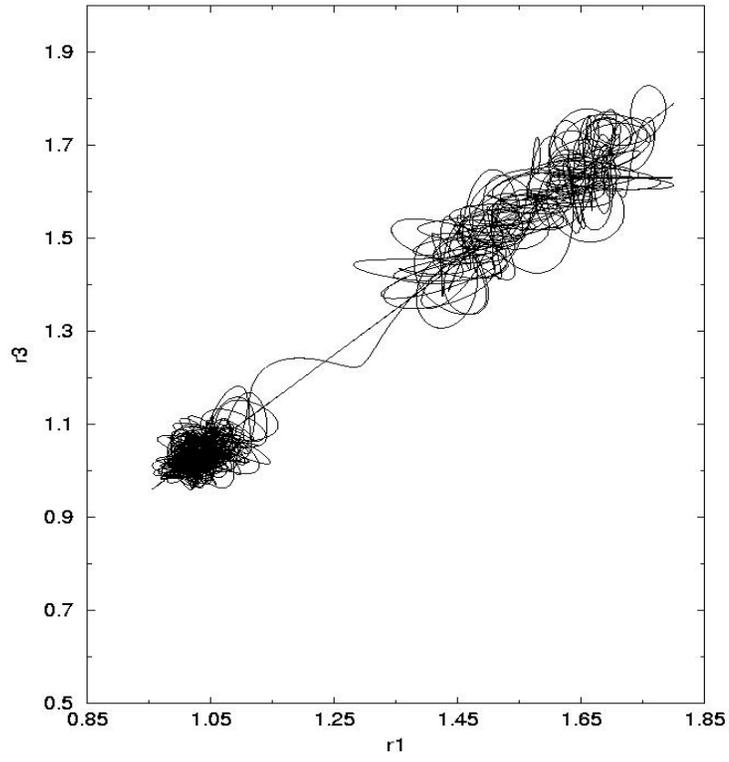

Fig. 2

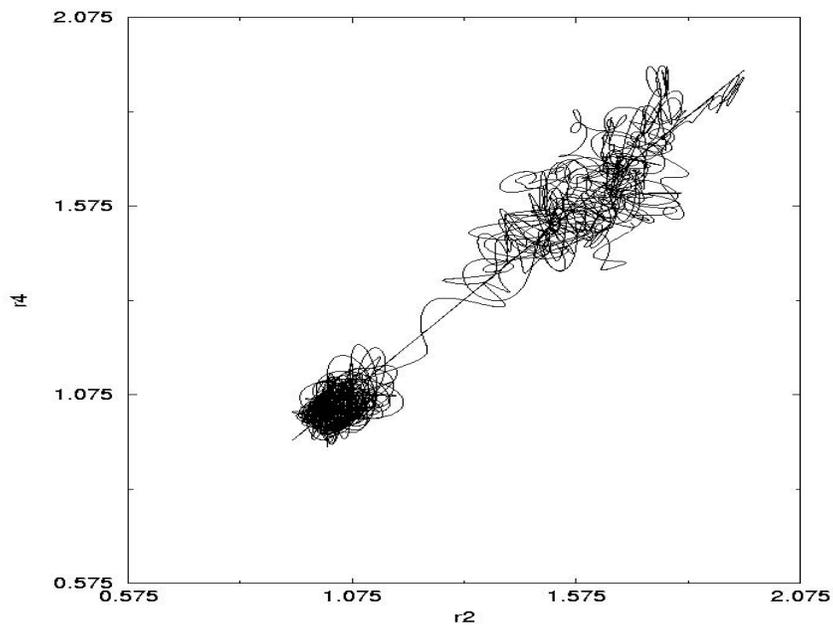

Fig. 3

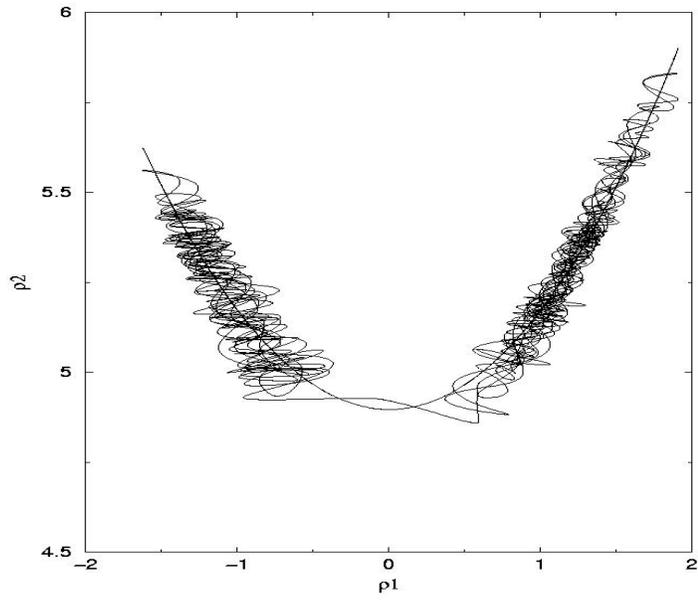

Fig. 4

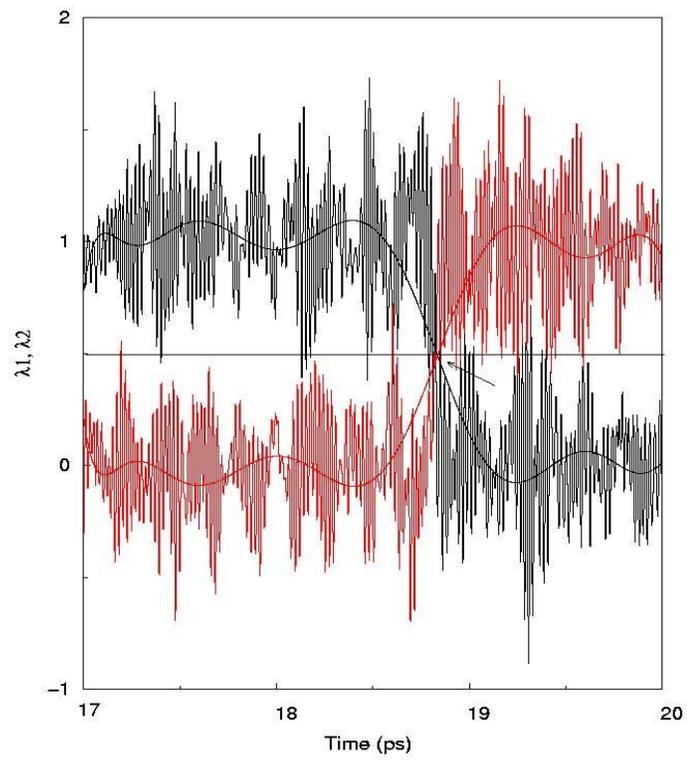

Fig. 5

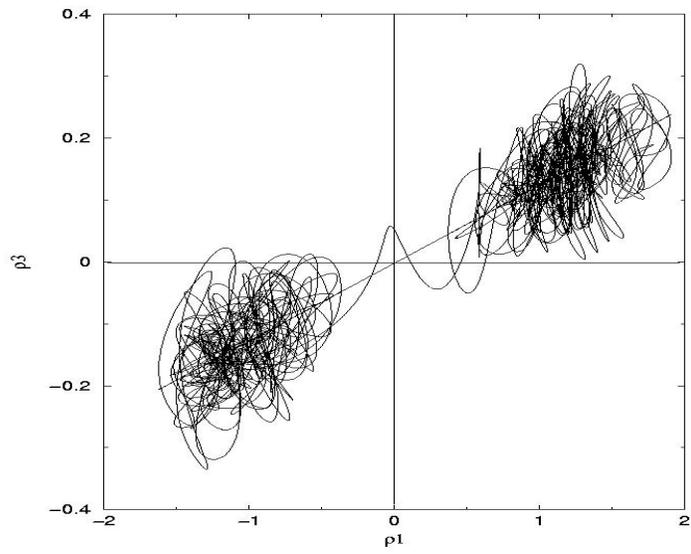

Fig. 6

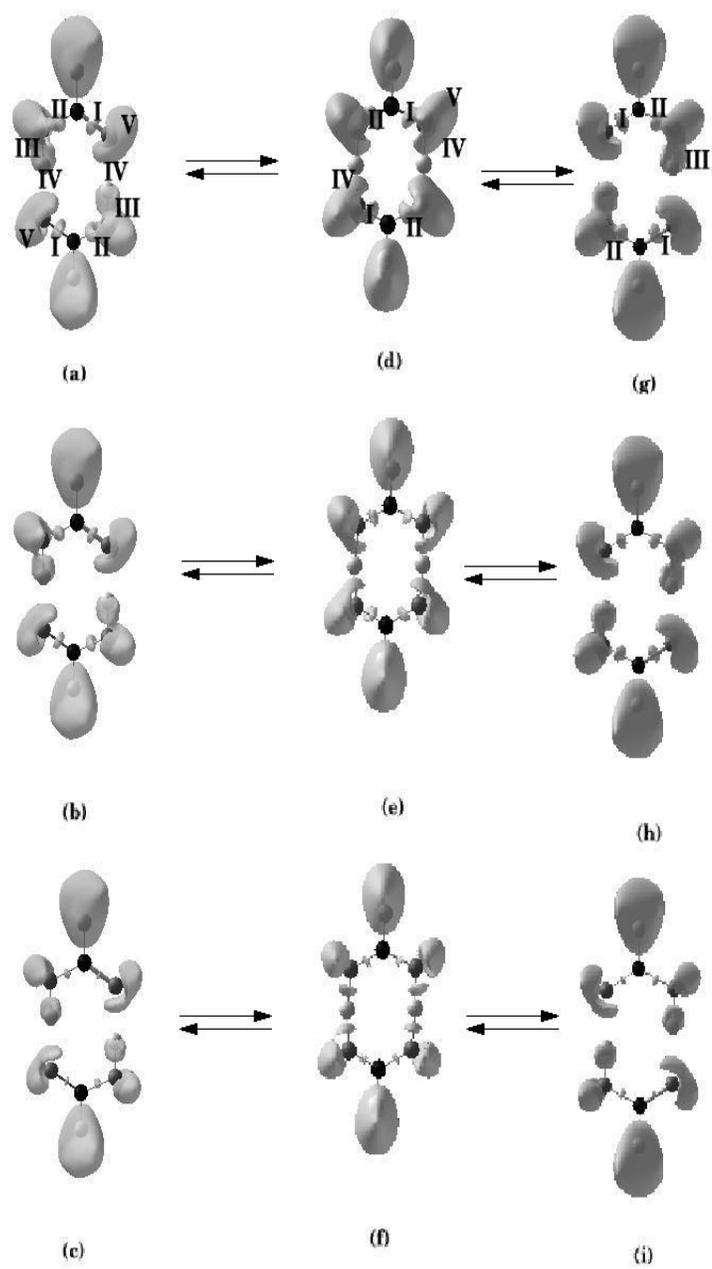

Fig. 7

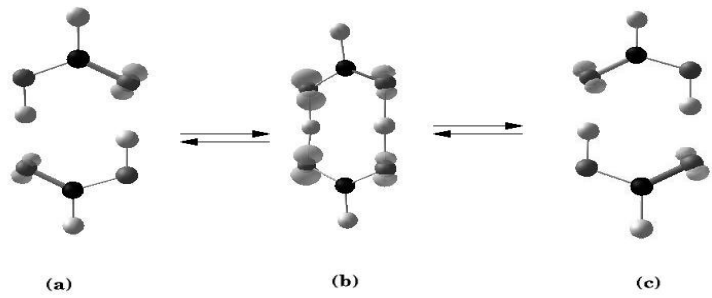

Fig. 8

# Table of Content

# A quantitative and a qualitative study of the resonance assisted double proton transfer in formic acid dimer.

**Sharan Shetty , Sourav Pal, Dilip G. Kanhere, Annick Goursot**